\documentclass{article}
\usepackage{float}
\usepackage[hidelinks]{hyperref} 
\usepackage{fancyhdr}

 \usepackage{graphicx} 

\usepackage[nonatbib,main,final]{neurips_2025}

\usepackage[backend=biber,style=ieee,maxnames=99]{biblatex}
\usepackage[utf8]{inputenc} 
\usepackage[T1]{fontenc}    
\usepackage{hyperref}       
\usepackage{url}            
\usepackage{booktabs}       
\usepackage{amsfonts}       
\usepackage{nicefrac}       
\usepackage{microtype}      
\usepackage{xcolor}         
\usepackage{booktabs}      
\usepackage{multirow}      
\usepackage{siunitx}       
\usepackage{makecell}      
\usepackage{comment}
\usepackage{algorithm}
\usepackage{algorithmic}
\usepackage{amsmath}
\usepackage{caption}
\usepackage{svg}
\usepackage{appendix,listings,xcolor}
\usepackage{float}  
\usepackage{fancyhdr}

\title{SoulX-Transcriber: A Robust End-to-End Framework for Multi-Speaker Speech Transcription}

\author{
\textbf{Yuhang Dai}\textsuperscript{1,2}\thanks{Equal contribution.} \and
\textbf{Haopeng Lin}\textsuperscript{2}\footnotemark[1] \and
\textbf{Zhennan Lin}\textsuperscript{1} \and
\textbf{Jiale Qian}\textsuperscript{2} \and
\textbf{Jun Wu}\textsuperscript{2} \and
\textbf{Hanke Xie}\textsuperscript{1,2} \and
\textbf{Hao Meng}\textsuperscript{2} \and
\textbf{Hanlin Wen}\textsuperscript{2} \and
\textbf{Chuang Ding}\textsuperscript{3} \and
\textbf{Shunshun Yin}\textsuperscript{2} \and
\textbf{Ming Tao}\textsuperscript{2} \and
\textbf{Lei Xie}\textsuperscript{1} \and
\textbf{Xinsheng Wang}\textsuperscript{2}\thanks{Corresponding author.
\\ 
\texttt{yhdai@mail.nwpu.edu.cn}, \texttt{linhaopeng@soulapp.cn},  \texttt{lxie@nwpu.edu.cn}, \\ \texttt{wangxinsheng@soulapp.cn}} \and \\
\textsuperscript{1}Audio, Speech and Language Processing Group (ASLP@NPU),\\
Northwestern Polytechnical University, Xi’an, China \\
\textsuperscript{2}Soul AI Lab, China \\
\textsuperscript{3}Moonstep AI, China \\
}

\begin{document}


\maketitle
\thispagestyle{fancy}

\vspace{-1.5em}

\begin{abstract}

\vspace{-1.em}

Recent advances in Automatic Speech Recognition (ASR) and Large Language Models (LLMs) have significantly improved speech understanding capabilities. However, multi-speaker speech transcription remains a challenging task, constrained by highly similar speaker voices, rapid turn-taking transitions and overlapping utterances. These challenges become particularly pronounced in real-world conversational audio, where speaker dynamics and acoustic conditions are highly variable.
This technical report presents SoulX-Transcriber, a unified multi-speaker transcription system that jointly models speaker diarization (SD) and ASR within an LLM-based framework. SoulX-Transcriber adopts a two-stage training strategy to improve both speaker discrimination and transcription robustness. In the first stage, speaker-aware multi-task continuous pre-training enhances speaker representation learning and boundary perception. In the second stage, supervised fine-tuning further optimizes the model for accurate end-to-end speaker-attributed transcription under complex multi-speaker conditions.
SoulX-Transcriber delivers strong performance and robustness across multiple public benchmarks, including AliMeeting, AISHELL-4, and AMI, while maintaining high adaptability to multi-domain scenarios.

\end{abstract}
\vspace{-1.em}

\hspace*{0.09\textwidth}
\begin{minipage}{0.9\textwidth}
\small
\textbf{Demo page:} \href{https://soul-ailab.github.io/soulx-transcriber/}{\textcolor{blue}{https://soul-ailab.github.io/soulx-transcriber}} \\
\textbf{Source code:} \href{https://github.com/Soul-AILab/SoulX-Transcriber}{\textcolor{blue}{https://github.com/Soul-AILab/SoulX-Transcriber}}
\end{minipage}

\begin{figure}[H]
    \centering
    \vspace{-0.3cm}
    \includegraphics[width=0.9\linewidth]{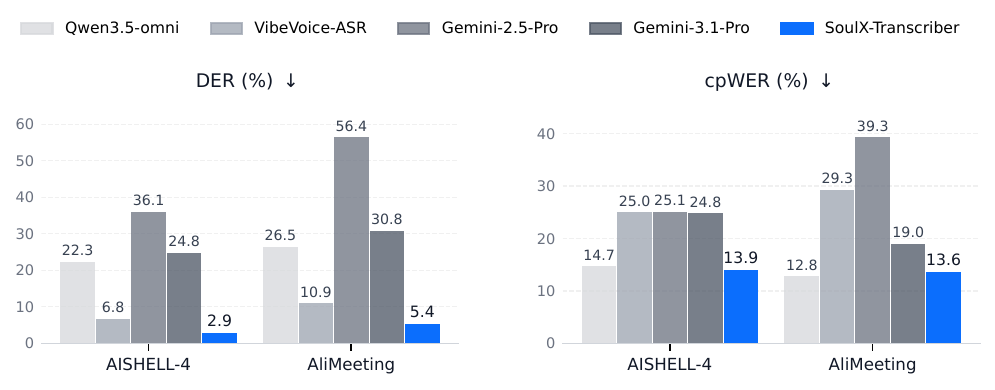}
    \caption{Performance of SoulX-Transcriber on AliMeeting and AISHELL-4.}
    \label{fig:soulxtranscriber}
\end{figure}

\vspace{-0.5cm}

\section{Introduction}
Natural human conversations feature complex multi-speaker behaviors. Against the rising industrial demands for multi-speaker ASR in meeting note generation and customer service auditing, the core objective of multi-speaker speech understanding is to resolve the critical question of \textit{“who spoke what and when”}. Yet real-world conversational audio poses great technical hurdles: frequent speaker alternations, widespread speech overlap, unstable acoustic environments, and pronounced voice similarity between speakers—especially those of the same gender. These artifacts severely hinder accurate speaker attribution and precise speaker boundary localization. To tackle these pain points, we unify Speaker Diarization (SD) and Automatic Speech Recognition (ASR) into a single end-to-end task, termed \textbf{SDR}.


Traditional SDR systems typically adopt cascaded pipelines composed of Voice Activity Detection (VAD), Speaker Verification (SV), and Automatic Speech Recognition (ASR)~\cite{ trainshort}. Although such modular architectures are interpretable and relatively flexible, they often suffer from significant error propagation and high system complexity. These limitations become more pronounced in conversational scenarios with intensive speaker interactions and overlapping speech~\cite{peng2026vibevoice}. Recent advances in Large Audio-Language Models (LALMs)~\cite{lalm_asr01, lalm_asr02} have introduced a new paradigm for end-to-end multi-speaker speech understanding. By jointly modeling acoustic signals and textual representations within a unified autoregressive framework, LALM-based systems can directly generate structured outputs containing speaker identities, timestamps, and transcriptions~\cite{lalm_sd01, lalm_sd02}. This unified modeling capability substantially simplifies system design while enabling stronger cross-task interaction between speaker modeling and speech recognition.



To improve speaker attribution capability within LALM-based SDR frameworks, existing studies mainly focus on several directions, including speaker-aware encoding and speaker enrollment mechanisms~\cite{tagspeech, yin2025speakerlm}, multimodal and long-context conversational modeling~\cite{3dspeaker, peng2026vibevoice}, hierarchical speaker classification and iterative speaker reasoning~\cite{glsc_sdr, lin2026speakerreasoner}, as well as post-processing refinement for diarization consistency and transcription readability~\cite{wang2024diarizationlm, post_process4sd}. These approaches improve multi-speaker transcription performance from different perspectives, such as enhancing speaker consistency, strengthening discrimination between acoustically similar speakers, and improving robustness under complex conversational conditions. 

Despite these advances, most existing methods primarily rely on architectural modifications or inference-stage optimization, while insufficiently addressing speaker representation learning during training. As a consequence, learned speaker representations often lack adequate intra-speaker compactness and inter-speaker discriminability, limiting robustness under challenging conversational conditions involving acoustically similar speakers, rapid turn transitions, and intensive speech overlap. These limitations significantly constrain the generalization capability of existing SDR systems in real-world conversational environments.

To address these challenges, we propose SoulX-Transcriber, an effective multi-speaker transcription system designed for robust conversational speech understanding. SoulX-Transcriber adopts a two-stage training framework consisting of speaker-aware multi-task continuous pre-training followed by SDR-oriented supervised fine-tuning. Without modifying the backbone architecture of the underlying LALM, the proposed framework explicitly enhances speaker representation learning and improves the model’s capability in speaker discrimination, speaker boundary perception, and overlapping speech recognition.


The main contributions of this work are summarized as follows:

\begin{itemize}
    \item \textbf{End-to-End Multi-Speaker Transcription Model}. We present SoulX-Transcriber, a unified SDR system capable of processing long-form conversational audio while directly generating structured outputs containing timestamps, speaker labels, and transcribed text.

    \item \textbf{Conversation-Oriented Simulation Data Pipeline}. We develop a scalable conversational data simulation pipeline that automatically retrieves acoustically and semantically suitable reference audio based on dialogue content, enabling the construction of more natural and contextually consistent multi-speaker training data.
    
    \item \textbf{Strong Performance on Multi-Speaker Transcription Tasks}. 
    SoulX-Transcriber achieves strong performance across multiple public benchmarks, including AliMeeting~\cite{alimeeting}, AISHELL-4~\cite{aishell4}, and AMI~\cite{ami}, while maintaining strong adaptability to real-world multi-domain conversational audio scenarios.    
    
\end{itemize}



\section{Method}
In this section, we introduce the overall framework of SoulX-Transcriber, including the complementary data engineering pipeline and the model training strategy. The data pipeline consists of large-scale pseudo-labeled conversational data and simulated multi-speaker dialogue data. We then describe the training and optimization process of the proposed SDR system.


\subsection{Data Processing}
High-quality large-scale multi-speaker conversational data is essential for training robust SDR systems. However, manually annotating speaker-attributed conversational audio is extremely expensive and difficult to scale, especially under complex conditions involving overlapping speech and rapid speaker transitions. To address this challenge, we construct a complementary data engineering framework consisting of two types of training data: pseudo-labeled real conversational data and simulated multi-speaker dialogue data.

The pseudo-labeled data preserves real-world acoustic characteristics and conversational dynamics, enabling the model to learn realistic speech distributions. However, its label quality is inherently constrained by the performance of automatic diarization and transcription systems, particularly for acoustically similar speakers. In contrast, simulated dialogue data provides controllable speaker diversity and conversational structures, which effectively improves speaker discrimination capability and out-of-domain generalization.


\subsubsection{Labeling Dialogue Data}

To construct large-scale conversational data, we build a cascaded pseudo-labeling pipeline on unlabeled multi-speaker audio recordings. The pipeline consists of three stages: speech activity detection, transcription generation, and speaker clustering.


\textbf{Speech Segmentation}. We first apply silero~\cite{Silero_VAD} and pyannote-vad~\cite{pyannote} to perform VAD. The outputs of the two VAD systems are aligned and merged to obtain more robust speech regions with improved boundary accuracy. Based on the detected speech regions, we further apply the pyannote speaker diarization pipeline~\cite{pyannote} to refine speaker turn boundaries and split the long-form audio into speaker-aware utterance-level segments.


\textbf{Multi-ASR Transcription}. Each speech segment is transcribed using multiple heterogeneous ASR models. We then perform hypothesis consensus fusion~\cite{dai2026wenetspeech} to obtain the final transcription result together with a confidence score. Segments with low transcription confidence are filtered out to improve pseudo-label reliability.


\textbf{Speaker Clustering}. For the retained speech segments, speaker embeddings are extracted and clustered using HDBSCAN~\cite{hdbscan} within each session. Segments assigned to the same cluster are treated as belonging to the same pseudo speaker identity. Neighboring segments with identical speaker labels and short temporal gaps are further merged into longer speaker turns.

\subsubsection{Multi-Speaker Dialogue Data simulation}
To further improve speaker discrimination capability and conversational diversity, we additionally construct a multi-speaker dialogue simulation pipeline based on long-form multi-speaker speech synthesis. Compared with pseudo-labeled conversational data, simulated dialogue data provides better controllability over speaker identity, dialogue structure, and conversational composition. It also enables scalable construction of difficult training samples involving acoustically similar speakers and complex multi-speaker interactions, which are difficult to obtain through automatic annotation pipelines alone.
The overall simulation pipeline consists of four stages: dialogue text construction, reference audio construction, speaker-reference matching, and dialogue audio generation.


\textbf{Dialogue Text Construction}. 
We collect large-scale conversational text data from podcasts, novels, and dialogue-centric corpora in both Chinese and English. An LLM-based text analysis module is then used to identify speaker roles and construct structured multi-speaker dialogue scripts. To maintain conversational coherence and interaction quality, the number of speakers in each dialogue sample is controlled between 3 and 8.

\begin{figure}[h]
    \centering
    \includegraphics[clip, trim=0cm 4cm 0cm 4cm, width=1.0\textwidth]{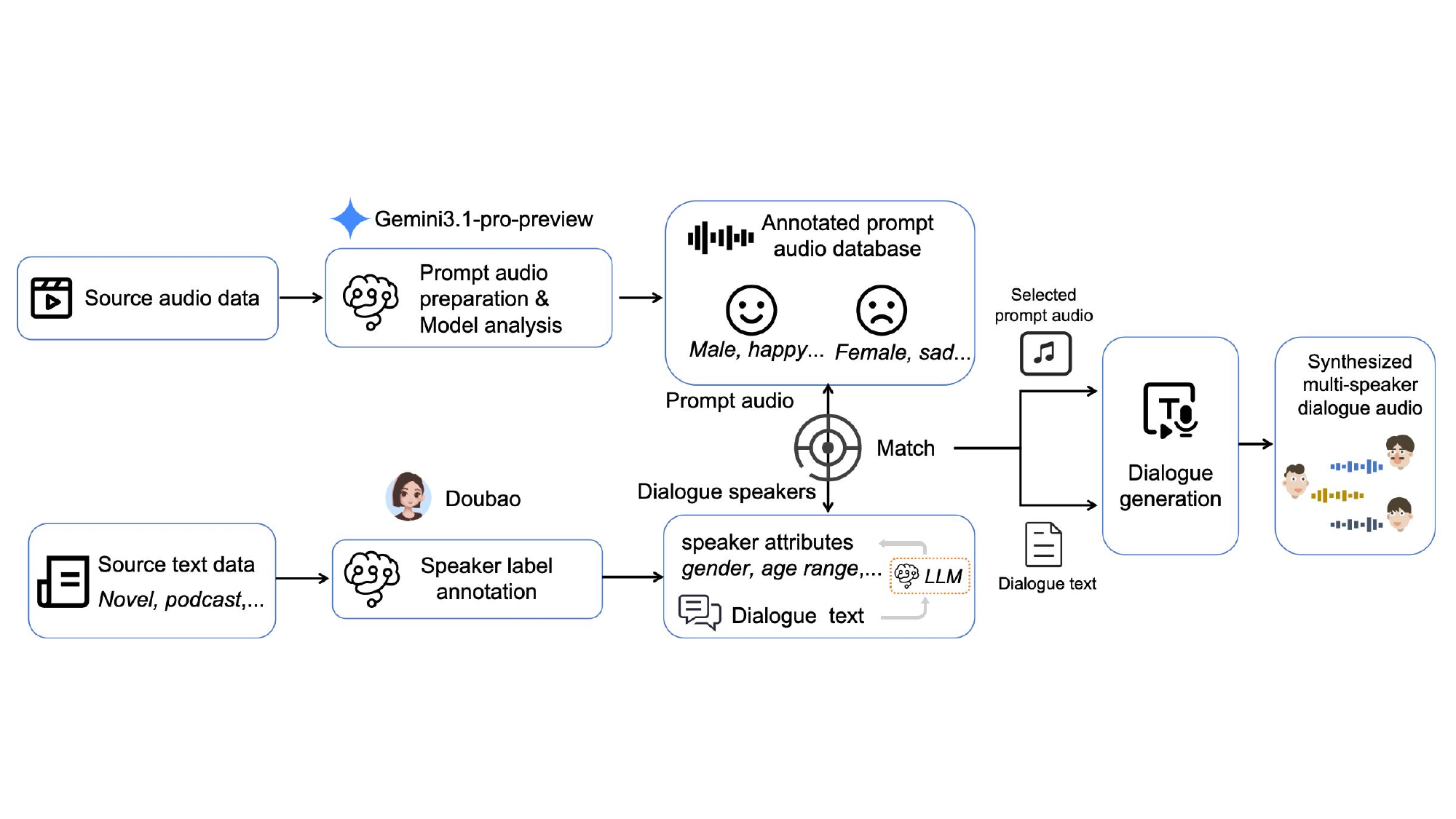}
    \caption{Pipeline for multi-speaker dialogue data simulation pipeline.}
    \label{fig:simulation_pipeline}
\end{figure}

\textbf{Reference Audio Construction}.
Reference audios are collected from long-form conversational recordings and multimedia content. After VAD-based segmentation, short speech clips with durations between 3 and 10 seconds are retained as candidate reference audios.

To ensure synthesis quality and speaker diversity, each reference audio is annotated with multiple speaker-related attributes, including gender, age range, emotion, speaking rate, pitch characteristics, timbre style, expression style, vocal characteristics, and speaking style. In addition, audio quality metrics such as UTMOS score\footnote{\url{https://github.com/fakerybakery/utmos}} and Signal-to-Noise Ratio (SNR) are used to maintain quality consistency across different reference speakers.

To support fine-grained speaker retrieval, we further construct a structured multi-dimensional speaker representation for each reference audio. Specifically, the textual labels of the nine speaker-related attributes are individually encoded using bge-m3~\cite{bge-m3}. The embedding vectors from all attribute dimensions are then stacked sequentially to form the speaker feature matrix of a reference audio:
\begin{equation}
    \mathbf{E_i} \in \mathbb{R}^{9 \times 1024}, i=1,...,N
\end{equation}
where each row corresponds to the embedding representation of one speaker attribute dimension. By stacking all reference audio feature matrices, we obtain the structured speaker representation database: $\mathbf{E} \in \mathbb{R}^{N \times 9 \times 1024}$.

\textbf{Speaker-Reference Matching}.
To generate conversationally consistent multi-speaker dialogues, we design a multi-dimensional speaker-reference matching mechanism based on attribute-wise semantic similarity. Given a target dialogue script containing $M$ speaker roles, the LLM first analyzes the speaker characteristics of each role and converts them into structured speaker attribute descriptions. Similar to the reference audio construction process, the attribute descriptions of each target speaker are encoded using bge-m3 and organized into a structured speaker feature matrix:
\begin{equation}
    \mathbf{Q_j} \in \mathbb{R}^{9 \times 1024}, j=1,...,M
\end{equation}
To measure the similarity between a target speaker and all candidate reference audios, we compute attribute-wise similarity scores between their corresponding embedding dimensions:
\begin{equation}
    \mathbf{V} = \mathbf{E} \cdot \mathbf{Q_j}^{\top} \in \mathbb{R}^{N \times 9 \times 9}
\end{equation}
Since each diagonal element represents the similarity between the same speaker attribute dimension, only the diagonal components are retained as valid attribute-wise similarity scores. This produces a 9-dimensional similarity vector for each reference audio. To further model the relative importance of different speaker attributes, we introduce a predefined attribute weight vector $\mathbf{w} = [w_1, w_2, \dots, w_9]^{\top} \in \mathbb{R}^{9}$. The final similarity score of each reference audio is computed as the weighted sum of the attribute-wise similarity scores:
\begin{equation}
    s_i = \mathbf{d}_i \cdot \mathbf{w} = \sum_{j=1}^{9} w_j \cdot \mathbf{d}_i^{(j)}
\end{equation}
Based on the weighted similarity scores, the top-k candidate reference audios are selected for each target speaker. 

Finally, an additional constraint-based filtering process is applied to ensure speaker diversity and synthesis consistency. Specifically, the selected reference audios are required to satisfy two constraints simultaneously: (i) different dialogue roles cannot use reference audios from the same original speaker; and (ii) the UTMOS score difference between matched reference audios must remain within a fixed threshold.

\subsection{SoulX-Transcriber}

\begin{figure}[t]
  \centering
  \vspace{-0.2cm}
  \makebox[\textwidth][c]{
    \includegraphics[clip, trim=0cm 1cm 0cm 1cm, width=1.0\textwidth]{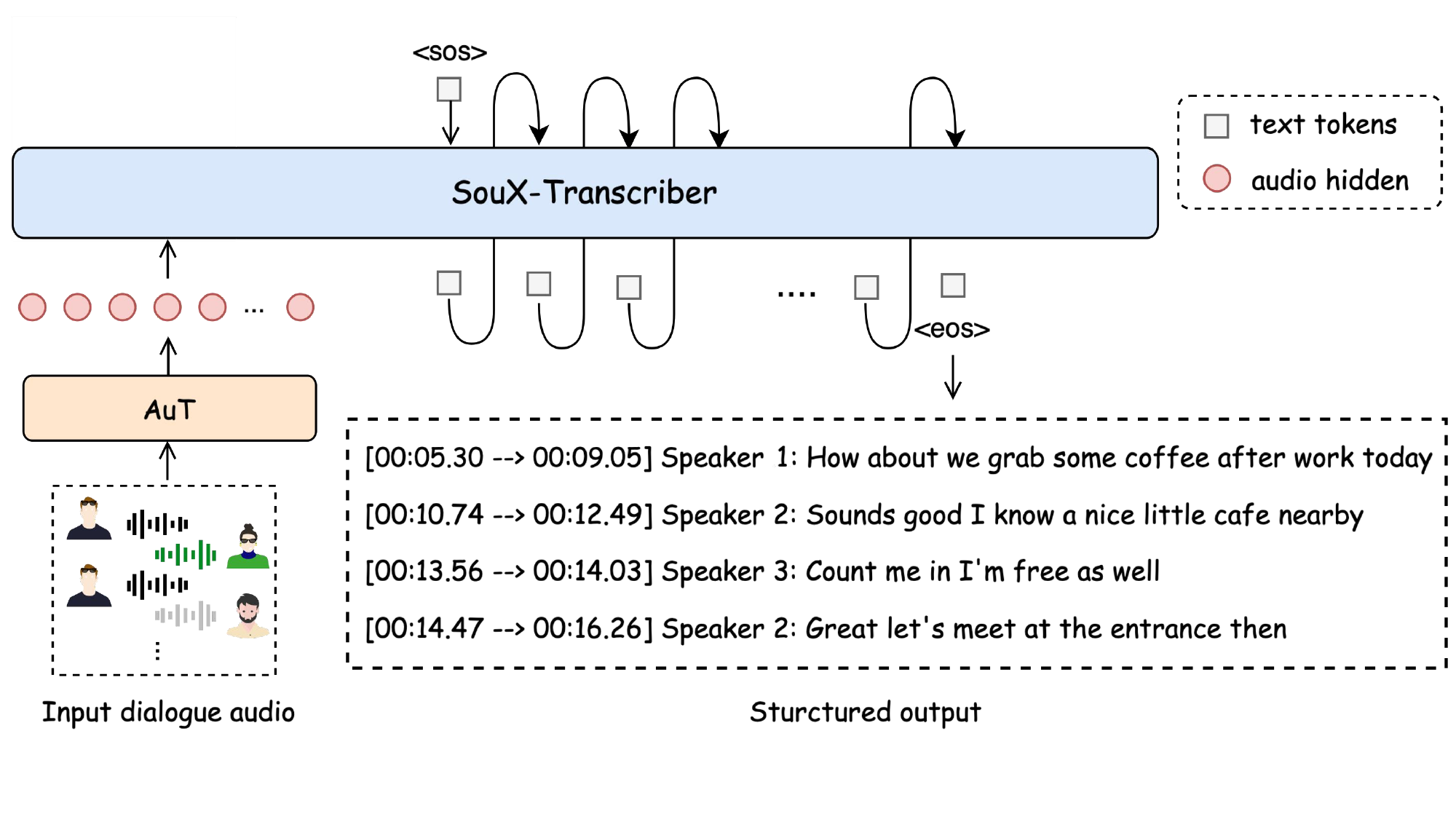}
  }
  \vspace{-0.6cm}
  \caption{The architecture of SoulX-Transcriber. SoulX-Transcriber accepts conversational audio clips with a maximum duration of 10 minutes and processes the input in a single forward pass to produce structured outputs containing timestamps, speaker assignments, and transcribed text.}
  \captionsetup[figure]{skip=1pt}
  \label{fig:train_stages}
  \vspace{-0.6cm}
\end{figure}


SoulX-Transcriber is built upon a large omni-modal framework Qwen3-Omni~\cite{qwen3_omni} with strong long-context audio understanding and autoregressive generation capabilities. Based on this backbone model, we introduce a two-stage speaker-aware training framework consisting of multi-task continuous pre-training followed by supervised fine-tuning. The structure of SoulX-Transcriber is illustrated in Figure~\ref{fig:train_stages}.


\subsubsection{Speaker-Aware Multi-Task Continuous Pre-training}

In the first training stage, we introduce a speaker-aware multi-task continuous pre-training framework to enhance the model’s capability in speaker representation learning, speaker turn perception, and multi-speaker conversational understanding. The training framework jointly optimizes multiple speaker-related tasks within a unified autoregressive generation paradigm.

\textbf{Speaker Turn Prediction (STP)}. The STP task is designed to improve the model’s temporal perception of speaker transition boundaries in conversational audio. During training, special boundary tokens are inserted into the target text sequence to explicitly model speaker turn transitions, enabling the model to better capture rapid turn-taking behaviors in multi-speaker conversations.


\textbf{Target Speaker Extraction and Recognition (TSER)}. The TSER task aims to strengthen speaker-conditioned speech understanding capability. Given a reference audio of a target speaker together with a multi-speaker conversational recording, the model is required to identify the speech segments belonging to the target speaker and generate the corresponding transcriptions. Timestamp supervision is additionally introduced to improve temporal localization accuracy and speaker-aware extraction capability.

\textbf{Speaker Verification (SV)}. To further enhance speaker discrimination capability~\cite{ren2025can}, we incorporate a SV task during continuous pre-training. Given two speech segments, the model predicts whether they belong to the same speaker. This task improves inter-speaker discriminability and strengthens robustness under acoustically similar speaker conditions.

\textbf{Speaker Diarization and Recognition (SDR)}. 
The SDR task serves as the core end-to-end multi-speaker transcription objective. Given a multi-speaker conversational audio segment, the model directly generates structured outputs containing speaker labels, timestamp boundaries, and transcribed text. This task jointly optimizes speaker attribution, temporal prediction, and speech recognition within a unified generation framework.

\textbf{Automatic Speech Recognition (ASR)}.
To preserve the general speech recognition capability of the backbone model, we additionally introduce a moderate amount of multilingual ASR data during continuous pre-training. Out-of-domain speech data is also incorporated to improve acoustic robustness and generalization capability across diverse recording conditions.

The training data ratio for the STP, TSER, SV, SDR and ASR tasks is approximately 2:2:1:5:1. The total training duration for continuous pre-training in the first stage reaches around 100,000 hours, among which roughly 3,000 hours consist of synthetic multi-speaker conversational data generated via our proposed simulation pipeline. We also use public datasets including AISHELL-4~\cite{aishell4}, AliMeeting~\cite{alimeeting}, AMI-SDM~\cite{ami} and English subset of MLC-SLM~\cite{mlc_slm}, while the remaining training data are internal proprietary corpus. All training audio samples are chunked into 5-minute segments, with a maximum segment length capped at 10 minutes.

\subsubsection{Supervised Fine-Tuning (SFT)}
Although the first-stage continuous pre-training substantially improves the model’s speaker-aware conversational understanding capability, the large-scale pseudo-labeled data inevitably introduces label noise and speaker attribution uncertainty, which limits the model’s robustness on high-precision SDR tasks.

To further improve speaker attribution accuracy, instruction consistency, and generalization capability, we perform a second-stage supervised fine-tuning process using high-quality annotated SDR data. The fine-tuning dataset consists of manually annotated conversational data together with carefully filtered simulated dialogue data, with a total duration of approximately 1,000 hours.

Through the two-stage training framework, SoulX-Transcriber progressively acquires speaker representation learning capability from large-scale conversational data and subsequently adapts to high-precision end-to-end SDR generation tasks under complex multi-speaker conversational conditions.

\begin{figure}[t]
    \centering
    \includegraphics[clip, trim=0cm 2.5cm 0cm 2.5cm,width=1.0\textwidth]{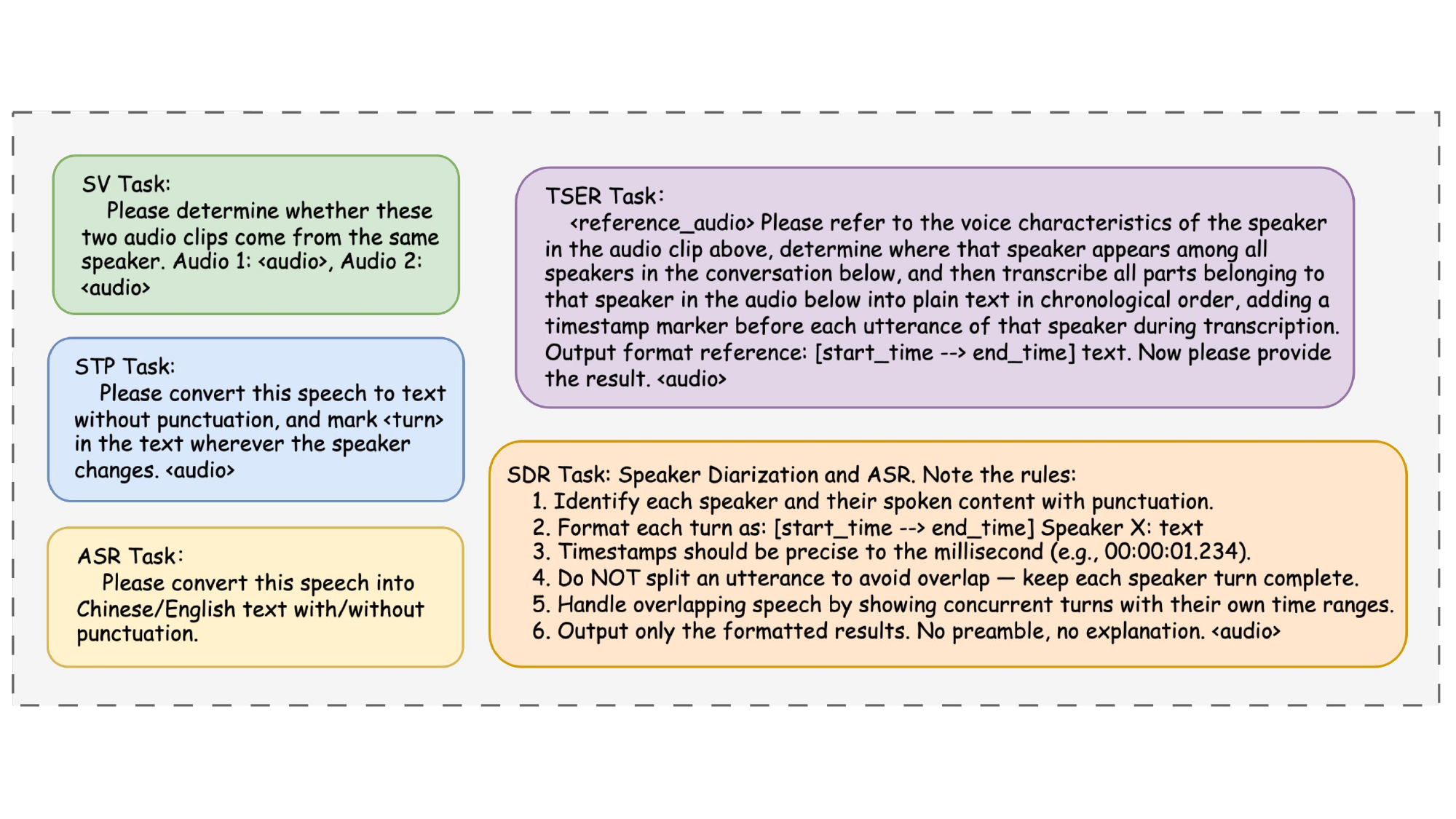} 
    \caption{Prompt for LLM used in Speaker-Aware Multi-Task continuous pre-training stage.}
    \label{fig:architecture}
\end{figure}

\section{Performance}

\subsection{Evaluation Dataset}

We evaluate SoulX-Transcriber on three widely used multi-speaker meeting benchmarks: AMI~\cite{ami}, AliMeeting~\cite{alimeeting}, and AISHELL-4~\cite{aishell4}. AMI is an English meeting corpus, while AliMeeting and AISHELL-4 are Mandarin far-field meeting datasets. Following common evaluation settings, we use the Single Distant Microphone (SDM) subset for AMI and the first channel of the 8-channel microphone array for AliMeeting and AISHELL-4. Following prior work~\cite{kanda2022transcribe, tagspeech}, a turn-group-based segmentation strategy is adopted, where consecutive speaker turns without silence gaps are merged into an evaluation segment.

To evaluate long-context speaker tracking capability, we additionally construct 5-minute evaluation sets on AliMeeting and AISHELL-4. Long-form recordings better expose speaker drift, long-range identity confusion, and accumulated diarization errors. Beyond public benchmarks, we further construct three internal manually annotated test sets covering daily conversations, movies, and podcasts. Each sample is approximately five minutes long. These internal benchmarks are designed to evaluate the generalization capability of SoulX-Transcriber under diverse open-domain multi-speaker scenarios beyond conventional meeting-style recordings.

\subsection{Evaluation Metrics}
The SDR task requires the system to generate accurate transcripts and timestamps, while correctly attributing each utterance to the corresponding speaker. To evaluate these capabilities, we adopt the following metrics.

\textbf{Diarization Error Rate (DER)}. 
DER measures speaker diarization performance by calculating the total duration of false alarms, missed speech, and speaker confusion relative to the reference speech duration.

\textbf{Word Error Rate (WER)}.
WER evaluates pure speech recognition performance by concatenating all utterances in chronological order while ignoring speaker identities.

\textbf{Concatenated minimum-Permutation Word Error Rate (cpWER)}.
cpWER jointly evaluates transcription quality and speaker attribution accuracy. It is computed by concatenating utterances belonging to the same speaker and searching for the speaker permutation with the minimum word error rate~\cite{cpWER_watanabe2020chime}.


\textbf{Speaker Attribution Gap ($\Delta cp$)}.
The difference between cpWER and WER reflects the additional errors introduced by incorrect speaker attribution. Therefore, $\Delta cp$ serves as an indicator of speaker diarization performance independent of transcription errors.

\subsection{Results}


\begin{table}[H]
\caption{The result on AliMeeting, AISHELL-4 and AMI-SDM. \textbf{Bold number} indicates the best result and \underline{underlined number} indicates the second-best result. Meanwhile, $\dagger$ indicates closed-source models.}
\label{table:short_result}
\renewcommand{\arraystretch}{1.3} 
\setlength{\tabcolsep}{0.1pt} 
\small 
\begin{tabular}{lcccccccccccc}
\hline
\multirow{2}{*}{\textbf{Models}} & \multicolumn{4}{c}{AISHELL-4}       & \multicolumn{4}{c}{AliMeeting}  & \multicolumn{4}{c}{AMI-SDM}   \\ \cline{2-13} 
    & DER↓  & WER↓  & cpWER↓ & $\triangle$cp↓  & DER↓  & WER↓ & cpWER↓ & $\triangle$cp↓ & DER↓  & WER↓  & cpWER↓ & $\triangle$cp↓ \\ \hline
Vibevoice-ASR~\cite{peng2026vibevoice} & \underline{6.77}  & 21.40  & 24.99 & 3.59 & \underline{10.92} & 27.40 & 29.33  & 1.93 & \underline{13.43} & \textbf{24.65} & \textbf{28.82}  & \underline{4.17} \\
Gemini-2.5-Pro$\dagger$ & 36.07 & 19.81 & 25.11 & 5.30 & 56.39          & 30.16 & 39.29  & 9.13 & 50.28 & 31.66 & 39.98  & 8.32 \\
Gemini-3.1-pro-preview$\dagger$ & 24.84 & 24.86 & 24.81 & -0.05 & 30.76  & 18.82 & 18.99  & \textbf{0.17} & 40.40 & 30.82 & 32.97 & \textbf{2.15} \\
Qwen3.5-omni$\dagger$ & 22.33 & \underline{15.13} & \underline{14.71} & \textbf{-0.42} & 26.46 & \textbf{12.44} & \textbf{12.79} & \underline{0.35} & 30.05 & 28.57 & 33.46 & 4.89 \\
SoulX-Transcriber  & \textbf{2.89}  & \textbf{14.16} & \textbf{13.90} & \underline{-0.26} & \textbf{5.39} & \underline{13.07} & \underline{13.61} & 0.54 & \textbf{11.67} & \underline{25.55} & \underline{32.78} & 7.23 \\ \hline
\end{tabular}
\end{table}

We follow the MeetEval~\footnote{\url{https://github.com/fgnt/meeteval}} evaluation protocol and report four metrics mentioned above. Table~\ref{table:short_result}, Table~\ref{table:long_result}, and Table~\ref{table:interal_testsets} present the performance of SoulX-Transcriber on public and internal multi-speaker conversational benchmarks, covering short-form, long-form, and more general-domain evaluation settings. 

\textbf{Short-form Benchmark.}
Table~\ref{table:short_result} reports the results on the short-form utterance-group benchmarks. SoulX-Transcriber achieves the best overall performance on AISHELL-4 and AliMeeting, consistently reducing DER, WER, and cpWER across all major metrics. In particular, the model maintains very small $\Delta cp$ values, indicating stable speaker attribution performance with limited additional speaker-related errors. On AMI-SDM, SoulX-Transcriber also achieves competitive performance. Notably, although the training data is primarily Mandarin-centric, the model still maintains reasonable performance on English conversational speech

\textbf{Long-form Benchmark.}
Table~\ref{table:long_result} shows the results on the 5-minute long-form benchmarks. Compared with existing open-source and commercial baselines, SoulX-Transcriber achieves substantially lower DER and cpCER on both AliMeeting and AISHELL-4. The model also maintains relatively small $\Delta cp$ values on long recordings, indicating stable speaker tracking and attribution capability over extended conversational contexts. 

\begin{table}[H]
\caption{Test results on the long-form audio (5 minutes) test sets AISHELL-4 and AliMeeting. $\dagger$ indicates closed-source models.}
\label{table:long_result}
\renewcommand{\arraystretch}{1.3} 
\setlength{\tabcolsep}{5pt} 
\small 
\centering
\begin{tabular}{lcccccccc}
\hline
\multirow{2}{*}{Model}          & \multicolumn{4}{c}{AliMeeting}& \multicolumn{4}{c}{AISHELL-4} \\ \cline{2-9} 
          & DER↓         & CER↓           & cpCER↓         & $\triangle$cp↓           & DER↓           & CER↓           & cpCER↓         & $\triangle$cp↓         \\ \hline
ViebVoice-ASR~\cite{peng2026vibevoice} & \underline{18}     & 29.72          & \underline{31.94}          & \underline{2.22}           & \underline{9.17}           & \underline{19.54}          & \underline{22.95}          & \underline{3.41}         \\
Gemini2.5-Pro$\dagger$          & 58.14        & 31.69          & 42.22          & 10.53          & 40.87          & 20.26          & 26.31          & 6.05         \\
Gemini-3.1-pro-preview$\dagger$ & 38.75        & \underline{26.75}          & 32.84          & 6.09           & 22.03          & 22.75          & 27.43          & 4.68         \\
SoulX-Transcriber      & \textbf{5.72} & \textbf{16.22} & \textbf{16.99} & \textbf{0.77}  & \textbf{7.73}  & \textbf{14.49} & \textbf{17.82} & \textbf{3.33}   \\ \hline
\end{tabular}
\end{table}

\textbf{General-Domain Benchmark.}
Table~\ref{table:interal_testsets} reports the results on the internal general-domain benchmarks, including daily conversations, movies, and podcasts. SoulX-Transcriber achieves strong overall performance across all domains, particularly on conversational and movie-style recordings with diverse acoustic environments and speaker interaction patterns. Despite the increased difficulty of podcast scenarios, the model still maintains competitive speaker attribution and transcription performance, demonstrating strong generalization capability under diverse multi-speaker conditions.

\begin{table}[H]
\caption{Test results on the internal benchmark, including Daily Conversation, Videos, and Podcast. All of them are about 5 minutes long. $\dagger$ indicates closed-source models.}
\label{table:interal_testsets}
\renewcommand{\arraystretch}{1.3} 
\setlength{\tabcolsep}{0.2pt} 
\small 
\centering
\begin{tabular}{lcccccccccccc}
\hline
\multicolumn{1}{l}{\multirow{2}{*}{Model}} & \multicolumn{4}{c}{Daily Conversaton}                        & \multicolumn{4}{c}{Movies}                                        & \multicolumn{4}{c}{Podcast}                                                               \\ \cline{2-13}
\multicolumn{1}{l}{} & DER↓ & WER↓ & cpWER↓ & $\triangle$cp↓ & DER↓ & WER↓ & cpWER↓ & $\triangle$cp↓           & DER↓ & WER↓ & cpWER↓ & $\triangle$cp↓ \\ \hline
Vibevoice-ASR~\cite{peng2026vibevoice} & \underline{2.76} & 30.34 & \underline{31.77} & \underline{1.43} & \underline{27.78} & 21.86 & 45.87 & 24.01 & \textbf{4.7} & \underline{8.88} & \textbf{14.58} & \underline{5.7} \\
Gemini-3.1-pro-preview$\dagger$ & 38.69 & \underline{29.14} & 36.72 & 7.58 & 34.87  & \underline{10.01}  & \underline{21.03} & \textbf{11.02} & 24.56 & 23.89 & 27.21 & \textbf{3.32}   \\
SoulX-Transcriber & \textbf{1.32} & \textbf{6.73} & \textbf{7.31} & \textbf{0.58} & \textbf{23.56} & \textbf{5.17} & \textbf{20.58} & \underline{15.41}    & \underline{21.15} & \textbf{7.5} & \underline{19.37} & 11.87 \\ \hline
\end{tabular}
\end{table}

Overall, the experimental results show that SoulX-Transcriber achieves robust and scalable multi-speaker conversational understanding capability across short-form, long-form, and general-domain scenarios, while maintaining strong speaker attribution accuracy and transcription quality under complex conversational conditions.

\section{Conclusions}

In this report, we present SoulX-Transcriber, an end-to-end multi-speaker transcription system designed for complex multi-speaker conversational scenarios. By combining a two-stage speaker-aware training framework with a scalable simulated dialogue data pipeline featuring structured speaker-attribute matching, SoulX-Transcriber achieves strong performance across diverse benchmarks while maintaining robust speaker attribution and transcription capability under challenging conversational conditions.

\printbibliography

\newpage




\end{document}